\documentclass[sigconf, nonacm]{acmart}
\acmJournal{PACMPL}
\acmVolume{1}
\acmNumber{SRC POPL} 
\acmArticle{1}
\acmYear{2019}
\acmMonth{1}
\acmDOI{} 
\startPage{1}

\setcopyright{none}

\usepackage{booktabs}   
\usepackage{subcaption} 

\usepackage{xparse}
\usepackage{amsthm}

\usepackage{LambdaLat}
\usepackage{LambdaLatSyntax}
\usepackage{LambdaLatNotations}
\usepackage{LambdaLatDynamicSemantics}
\usepackage{LambdaLatStaticSemantics}
\usepackage{LambdaLatArithmetic}

\usepackage[USenglish]{babel}
\usepackage[T1]{fontenc}
\usepackage[utf8]{inputenc}
\usepackage[scaled]{beramono}
\usepackage{enumitem}
\usepackage{listings}
\usepackage{upquote}
\usepackage{booktabs}
\usepackage{setspace}
\usepackage{xspace}
\usepackage{subcaption}
\usepackage{mathpartir}
\usepackage{balance}

\theoremstyle{acmdefinition}

\theoremstyle{acmplain}

\SetTracking[spacing = {25*, 166, }]{ encoding = *, family = fvm }{-40}
\SetTracking[spacing = {25*, 166, }]{ encoding = *, shape = {sc, si, scit} }{-30}

\lstdefinelanguage{Scala}{
  morekeywords={
    abstract,case,catch,class,def,
    do,else,extends,false,final,finally,
    for,if,implicit,import,lazy,match,
    new,null,object,override,package,
    private,protected,return,sealed,
    super,this,throw,trait,true,try,
    type,val,var,while,with,yield},
  morekeywords=[2]{
    asLocal,asLocalFromAll,asLocalFromAllSeq,
    call,capture,connect,connected,egg,from,listen,
    localOn,on,placed,remote,sbj,sharedOn,to,
    Multiple,Optional,Single,Tie,withLatency},
  sensitive=true,
  morecomment=[l]{//},
  morecomment=[n]{/*}{*/},
  morecomment=[s][identifierstyle]{`}{`},
  morestring=[b]",
  morestring=[b]',
  morestring=[b]"""
}

\begin{document}

\title{A Core Calculus for Static Latency Tracking\\ with Placement Types}
\subtitle{Student Research Competition @ POPL 2019}

\author{Tobias Reinhard}
\orcid{0000-0003-1048-8735}             
\affiliation{
  \institution{Technische Universit\"at Darmstadt}
	\position{Graduate Student}
}

\begin{abstract}

Developing efficient geo-distributed applications is challenging as 
programmers can easily introduce computations that entail high latency communication. 
We propose a language design which makes latency explicit and extracts type-level bounds for a computation's runtime latency. We present our initial steps with a core calculus that enables extracting provably correct latency bounds and outline future work.

\end{abstract}

\maketitle


\NewDocumentCommand{\nextTerm}{}{\ensuremath{\term^\prime}\xspace}
\NewDocumentCommand{\nextLat}{}{\ensuremath{\lat^\prime}\xspace}
\NewDocumentCommand{\op}{}{\ensuremath{\p^\prime}\xspace}
\NewDocumentCommand{\oPeer}{}{\ensuremath{\Peer^\prime}\xspace}
\NewDocumentCommand{\iSet}{}{\ensuremath{\pSet[\Peer]}\xspace}
\NewDocumentCommand{\oISet}{}{\ensuremath{\pSet[\oPeer]}\xspace}

\NewDocumentCommand{\tOp}{}{\ensuremath{\term_{\op}}\xspace}
\NewDocumentCommand{\latCon}{}{\ensuremath{\lat_c}\xspace}
\NewDocumentCommand{\belFctTerm}{}{\ensuremath{\hat{\fctTerm}}\xspace}
\NewDocumentCommand{\latRun}{}{\ensuremath{\lat_R}}
\NewDocumentCommand{\latType}{}{\ensuremath{\lat_T}}

\NewDocumentCommand{\sizedType}{m m}{\ensuremath{(#1, \eqClassT{#2})}}
\NewDocumentCommand{\fullType}{m m m}{\ensuremath{(#1, \eqClassT{#2}, \eqClassT{#3})}}

\NewDocumentCommand{\mutualTieRelation}{}{\ensuremath{\leftrightarrow}}
\NewDocumentCommand{\peersAreTied}{m m}{\ensuremath{#1 \mutualTieRelation #2}}

\NewDocumentCommand{\sizeArg}{}{\ensuremath{\size_\argTerm}\xspace}
\NewDocumentCommand{\latArg}{}{\ensuremath{\lat_\argTerm}\xspace}
\NewDocumentCommand{\nextArg}{}{\ensuremath{\argTerm^\prime}}
\NewDocumentCommand{\sizeNextArg}{}{\ensuremath{\size_{\nextArg}}}
\NewDocumentCommand{\nextSize}{}{\ensuremath{\size^\prime}\xspace}

\NewDocumentCommand{\oFctTerm}{}{\ensuremath{g}\xspace}

\NewDocumentCommand{\Bfun}{}{\ensuremath{B^\rightarrow}}

\newtheorem{myTheo}{Theorem}

\section{Introduction}
\label{sec:introduction}
	Developing efficient geo-distributed applications remains a challenging task. Efficiency is largely determined by latency caused by remote communication. Avoiding high-latency remote communication and exploiting locality is therefore imperative~\cite{Wittie:2010:ELI:1921168.1921201}. Distributed components are, however, often interconnected and local computations can trigger a chain of events causing high-latency remote computations~\cite{Weisenburger:2017:QRA:3105503.3105530,Luthra:2018:TAD:3210284.3210292,8354906}. Determining which local computations eventually lead to latency, introducing remote communication, often requires a global view which hinders modular development of geo-distributed software.
Also, the exact location where a remote computation is placed, matters. Communication among servers in a single data center, for instance, is much faster (under 2ms) than communication between geo-distributed data centres possibly located on different continents (over 100ms)~\cite{Bernstein:2017:GAS:3152284.3133931}.
	
	We build on the idea of making locations and latency explicit~\cite{StatLatTrackPlaceTypes} and adopt the approach that a computation's location and its entailed latency become part of its type. The type system can infer an upper bound on a computation's actual latency and can reject code containing wrong assumptions, e.g., on the latency caused by a method invocation. As a method signature already describes the latency its invocation entails, no global view is required anymore and code becomes more modular.

In this work, we present our work on formalizing the type system in a core calculus \langName, on extracting latency bounds, and on proving their correctness. We outline ongoing work about further extending the formalization and about enabling latency-saving refactorings. Finally, we discuss how we plan to evaluate this research line.

\vspace{1mm}
\noindent
{\bf Supervisors}: Guido Salvaneschi (Technische Universit\"at Darmstadt), Pascal Weisenburger (Technische Universit\"at Darmstadt).

\section{A Calculus for Latency}
\label{sec:latency-tracking}
	In the following we present \langName as well as the ideas behind its correctness proof. The goal of \langName is to track the location (called a \emph{peer}), on which a computation is run, and the latency it causes in the type system. In this calculus, a computation's latency refers to the weighted number of remote messages sent during the computation.
	The time a message needs to reach its recipient depends on the involved peers. For a fixed set of peers, e.g., found in geo-distributed data centers, we can assume fixed locations, and thus, known latency values, e.g., known from monitoring.
	We therefore use a function 
	$\LatContext : \PeerTypes \times \PeerTypes \rightarrow \Nats$
	to assign weights (i.e., latency approximation) \LatLookup{\Peer}{\oPeer} to the messages sent from peer \Peer to \oPeer.

	\paragraph{Dynamic Semantics}
		\langName is an extension of the typed lambda calculus~\cite{barendregt1984studies} where every computation is placed on a specified location.
		Peer types \Peer and peer instances \p specify type-level and runtime locations, respectively.
		Types are augmented by sizes and latency bounds. We use a fragment of Heyting arithmetic~\cite{Kohlenbach2008AppliedPT}  containing 0, \SucSymb, +, \minArithSymb and $\cdot$ to define those and prove arithmetic properties.
	
		The small step reduction relation \localEvalStepArrow{\pSet} describes a reduction step on a set of peer instances \pSet. 
		Locations and latency are explicit in the reduction semantics.
		Every intermediate result as well as the end result of a term's reduction is annotated by its location and the latency its reduction has caused. A peer evaluation context \peerEvalContext{\pSet}{\term}{\lat} describes a term \term to be evaluated on a set of peer instances \pSet where \lat is the latency that has been caused during the reduction so far.
		A reduction step \localEvalStepCont{\pSet}{\term}{\lat}{\nextTerm}{\nextLat} expresses that term \term is reduced on the peer instances \pSet to \nextTerm and that the tracked latency increases from \lat to \nextLat. Local reduction steps are standard and leave the tracked latency unchanged. Also, no latency-decreasing steps exist. Hence, we have $\lat \leq \nextLat$.
		However, every reduction step involving a message sent from a peer \Peer to \oPeer increases the tracked runtime latency by the weight \LatLookup{\Peer}{\oPeer}

		Consider the evaluation of the term \get{\op}{\synVal} on a set \iSet of \Peer-instances. The expression requests a value \synVal from \op and can be reduced by sending the request to \op, increasing the latency by \LatLookup{\Peer}{\oPeer}.
		Hence, the reduction step is \\
        \vspace{-0.34cm}
		$$
        \localEvalStepCont
			{\iSet}
			{\get{\op}
				 {\synVal}
			}
			{\lat}
			{\peerEvalContext{\singleton{\op}}{\synVal}{0}}
			{\lat + \LatLookup{\Peer}{\oPeer}}.
        $$
		A remote evaluation \peerEvalContext{\singleton{\op}}{\synVal}{0} starts with a remote latency of 0.
		When the result is transmitted, the remote latency is added to the local latency.
		Since we assume \synVal to be a value, it cannot be reduced any further and the next step is sending \synVal from \op to \iSet.
		Runtime latency thereby increases to $\lat + \LatLookup{\Peer}{\oPeer} + \LatLookup{\oPeer}{\Peer}$.

	\paragraph{Static Semantics}
		Assigning types to well-formed terms \term ensures that there is a sequence of reduction steps towards a value \peerEvalContext{\pSet}{\synVal}{\lat} and that \synVal belongs to that type. Our approach lifts the latency of every reduction step to the type level. Additionally we ensure termination of recursive functions by employing sized types~\cite{Abel2007TypebasedTA} and only allowing size-decreasing recursion.
		A type is a triple \fullType{\B}{\size}{\lat}. \B is a basic type like \Unit determining the kind of value a term reduces to, 
		\size and \lat are arithmetic terms representing the value's size and an upper bound on the latency caused during reduction. \eqClassT{\size} and \eqClassT{\lat} denote the equivalence classes of terms provably equal to \size and \lat, respectively, in Heyting arithmetic.

	\paragraph{Type-level Latency}
		Considering the term \get{\op}{\term} (similiar to the example above but without assuming \op and \term to be values) and abstracting over concrete peer instances: Evaluating this term means (i)~evaluating \op on the current peer \Peer, (ii)~sending a request for \term to the remote peer \oPeer, (iii)~waiting for its evaluation and (iv)~\oPeer sending the result to \Peer. 
		We lift the runtime latency
		$\lat_{\op} + \LatLookup{\Peer}{\oPeer} + \lat_\term + \LatLookup{\oPeer}{\Peer}$
		to the type level, as typing rule \TGetName shows:
		\vspace{-0.15cm}
		{\footnotesize \TGet}
	
	\paragraph{Latency Bounds}
		Analyzing a term's structure is not always sufficient to compute its exact runtime latency. In general, terms can have multiple reduction sequences resulting in different runtime latencies.
		We therefore consider type-level latency as an upper bound on all possible runtime latencies, the typed term can reduce to.
		For instance, considering the term \ifTerm{\term_c}{\term_t}{\term_f}: In any case, the condition $\term_c$ is evaluated and depending on the result also one of the branches $\term_t$ and $\term_f$. As shown by rule \TIfName, we extract an upper bound by taking the maximum over both branches' latency:
		\vspace{-0.15cm}
		{\footnotesize \TIf}

	\paragraph{Size-dependent Functions}
		In the case of functions, the latency can depend on the input's size. Considering a list processing function \fctTerm, where the processing of each element involves some latency \lat: For any list \argTerm of size \sizeArg the latency of an application \localApp{\fctTerm}{\argTerm} is $\sizeArg \cdot \lat$.
		In \langName, such a function's basic type has the form 
		$\forallNats
			{\size}
			{\ArrowNewDetails{\B}{\size}{\B^\prime}{\size^\prime}{\lat^\prime}}$. 
		It expresses that the function can handle arguments of type \B and arbitrary size \size and that every such argument is mapped to a value of type \fullType{\B^\prime}{\size^\prime}{\lat^\prime} where variable \size may occur free in $\size^\prime, \lat^\prime$.
		For instance, in the previous example, we get $\lat^\prime = \size \cdot \lat$.

	\paragraph{Size-decreasing recursion}		
		Recursion is a convenient way to define size-dependent functions. In \langName, the only way to express recursion is via a fixpoint operator. Our type system ensures that for every application to an argument of size \size, the recursive step is taken on a smaller argument of size $\nextSize < \size$. Since sizes are finite, our type system ensures termination. Thus, application of the fixpoint operator preserves the correctness of extracted latency bounds.

	\paragraph{Correctness}
		We have shown how our type system lifts runtime latency to the type level and extracts upper bounds for branching terms. We also showed how we can extract latency bounds for recursive function applications. Hence, we can prove the following:

		\begin{myTheo}[Correctness of Type-level Latency Bounds]
			Let \PlacementTypingEnv and \LocalTypingEnv be typing environments for placed and local variables, respectively. Let \ArithAssumpSet be a set of arithmetic assumptions and \Peer a peer type, \pSet a set of peer instances, \term a term, \synVal a value. Further, let \B be a basic type, \size a size and $\latRun, \latType$ latencies.
			Suppose 
			\hasType{\term}{\B}{\size}{\latType}
			and that there exists a reduction sequence for \term to a value	\peerEvalContext{\pSet}{\synVal}{\latRun}.
			Then $\latRun \leq \latType$ holds.
		\end{myTheo}

\section{Outlook}
	We are currently investigating whether the extracted type-level bounds are optimal regarding the worst-case runtime latency.
	This is particularly interesting for recursive functions where we need to check that an input exists that (i)~causes the estimated maximal number of recursive steps and (ii)~in each step meets the extracted latency bound.
		As message delay in distributed systems is non-deterministic, we plan to refine our approach by using probability distributions for the latency weights \LatLookup{\Peer}{\oPeer} instead of natural numbers.
	
	An important aspect to consider is to complement the (static) analysis provided by the type system with 
	actual latency measurements collected via monitoring. We believe that the combination of both can 
	provide correct feedback to the developers. To this end, we are working on a monitoring system 
	that provides realistic estimations for latency and retrofits them in the type system using methods 
	from continuous integration.

	We are currently implementing a prototype of the language presented in~\cite{StatLatTrackPlaceTypes} based on the type system of \langName. 	Eventually, we are going to implement type-based latency tracking in ScalaLoci~\cite{Weisenburger:2018:DSD:3288538.3276499}, 
	a multitier language whose type system keeps track of a computation's location similar to \langName. 
	
		Using ScalaLoci's extended type system, we are going to explore latency-saving refactorings. High-latency inducing computations often contain unnecessary remote communication. Relocating parts of the computation and only transmitting as few data as necessary helps to reduce latency. We believe that the combination of static location and latency information is sufficient to implement such refactorings.
	
	We plan to evaluate the type system's usability with controlled experiments and case studies on applications involving multiple geo-distributed data centres. Using platforms like Amazon AWS, we plan to use real locations for the data centers~\cite{GoogleDataCenterFAQ} and to specify realistic latency weights \LatLookup{\Peer}{\oPeer} for the connections.

\section{Related Work}
	This paper builds on our previous work presenting the design of a language which makes latency transparent to the programmer~\cite{StatLatTrackPlaceTypes}.
	To the best of our knowledge, no previous work formally explores type-level latency tracking to promote low-latency computations.
    
	\citet{jost2009worst} augment a type system with cost values to extract upper bounds on the worst-case execution time and heap space usage. 
	Their approach, however, targets embedded systems where both time and space bounds are important.
	
	\citet{DBLP:conf/rtss/DelangeF14} propose an incremental, model-based approach to analyse the validity of latency requirements in cyber-physical systems.
	 
	\citet{Cohen:2012:ET:2384616.2384676} present a type system raising the developer's awareness for inefficient code in terms of energy consumption. Their approach augments types by energy consumption patterns and uses type inference to track a program's energy consumption.
	
	Session types (e.g., \citet{Hu:2008:SDP:1428508.1428538}) have been successfully applied to distributed
	programming to check distributed protocols, but focus on protocol correctness rather than communication cost.

\bibliography{bibliography}

\end{document}